# An Analysis of Beamed Wireless Power Transfer in the Fresnel Zone Using a Dynamic, Metasurface Aperture

David R. Smith[1a)], Vinay R. Gowda[1], Okan Yurduseven[1], Stéphane Larouche[1], Guy Lipworth[2], Yaroslav Urzhumov[1], and Matthew S. Reynolds[3]

[1]Duke University, Department of Electrical and Computer Engineering, Center for Metamaterials and Integrated Plasmonics, Durham NC, 27708, USA.

[2]Intellectual Ventures, Metamaterials Commercialization Center, Bellevue, Washington, 98004, USA.

[3]University of Washington, Department of Electrical Engineering, Department of Computer Science & Engineering, Seattle, Washington, 98195, USA.

Wireless power transfer (WPT) has been an active topic of research, with a number of WPT schemes implemented in the near-field (coupling) and far-field (radiation) regimes. Here, we consider a beamed WPT scheme based on a dynamically reconfigurable source aperture transferring power to receiving devices within the Fresnel (near-zone) region. In this context, the dynamic aperture resembles a reconfigurable lens capable of focusing power to a well-defined spot, whose dimension can be related to a point spread function (PSF). Near-zone focusing can be achieved by generating different amplitude or phase profiles over the aperture, which can be realized using traditional architectures, such as phased arrays. Alternatively, metasurface guided-wave apertures can achieve dynamic focusing, with potentially lower cost implementations. We present an initial tradeoff analysis of the near-zone WPT concept, relating key parameters such as spot size, aperture size, wavelength, focal distance, and availability of sources. We find that approximate design formulas derived from the Gaussian optics approximation provide useful estimates of system performance, including transfer efficiency and coverage volume. The accuracy of these formulas is confirmed using numerical calculations.

**I. INTRODUCTION**

Despite the dramatic growth of wireless technology in the communications domain, the use of wireless technology to provide power to devices remains in its infancy, due to both technical as well as market related concerns. Wireless power transfer (WPT) can allow power to be delivered without requiring a wiring infrastructure—a useful feature especially for remote, difficult to access devices, as well as embedded devices and sensors. The challenge for WPT, however, is in achieving a high-efficiency system at reasonable transfer distances. The dominant approach to date to WPT has made use of the magnetic near-fields, in which power is transferred between source and receiver coils that are coupled through non-radiating magnetic fields at very low frequencies of operation (kilohertz through megahertz, for example)[1,2,3]. Because near-field magnetic WPT systems are safe in terms of human exposure and can be highly efficient at short distances, they have led to numerous commercialization efforts[4]. However, because the near-field coupling falls off rapidly with distance between

---

[a)] Author to whom correspondence should be addressed. Electronic mail: drsmith@duke.edu.

source and receiver (as theسسsixth power of the inverse distance)[5], near-field WPT schemes require the receiving device to be in close proximity to the power source. While this proximity constraint is less problematic for some applications, such as vehicle charging, it remains an inconvenience in other contexts, and can rule out entire application areas such as powering remote sensors at long ranges.

At the other extreme, power transfer can be accomplished using short wavelength radio frequency (RF) power radiated from a source aperture to a receive antenna or aperture[6,7]. The advantage of such a WPT system is that power can be transferred over very long distances to targets at arbitrary positions, in hard to access regions or embedded in other materials. The disadvantage for far-field systems is that the beam width from an aperture is limited by diffraction, with the result being that only a minute fraction of power supplied by the source is captured by the receive aperture. In far-field scenarios, for which the distance between source and receive apertures is greater than $d = 2D^2/\lambda_0$ (where D is the aperture dimension)[8], the ratio of the power captured by the receiver to the supplied power is governed by the Friis equation. To achieve even modest efficiency levels for WPT schemes in the far-field regime, enormous apertures would be required. In addition, if the target to be powered is in motion, such as might be the case for an unmanned aerial vehicle (UAV) or autonomous automobile, then the source aperture would need to either be mechanically scanned or electronically reconfigurable.

If the distance between source and receiver is within the Fresnel zone (also termed the "radiating near-field," $d < 2D^2/\lambda_0$), and a line of sight is available, then a high efficiency WPT system can be realized by using a large aperture that acts as a lens, concentrating electromagnetic energy at a focal point where the receiving aperture is positioned[9,10,11]. In this scenario, a method of dynamically creating and moving a focal spot is needed. Recent intense research and development in the area of metasurface apertures—guided wave structures that radiate energy through an array of patterned irises—has shown a path to extremely low-cost, mass-producible reconfigurable apertures that could be configured for the WPT application[12,13]. The metasurface antenna is a passive device, in the sense that active phase shifters and amplifiers are not required to achieve the desired element tuning. Thus, for many implementations there is minimal power dissipation in beam steering, in contrast to typical phased array or electronically scanned antenna systems. Through the control over the phase or amplitude of each radiating element, holographic patterns can be created on the metasurface that mimic the functionality of Fresnel lenses or other diffractive optical elements. The metasurface aperture thus effectively can function as a low-cost, dynamically reconfigurable lens that consumes minimal power. A Fresnel zone WPT system based on a metasurface aperture can potentially achieve very high efficiency at minimal cost.

There are a variety of factors that must be considered for achieving a viable Fresnel zone WPT platform. In particular, the wavelength of operation represents a critical design choice. Ideally, short wavelength radiation is desirable, since very small focal spot sizes can be formed with moderate sized apertures. However, the cost of microwave sources increases dramatically at shorter wavelengths, forming a crucial tradeoff decision for the system. Currently, fairly large, dynamically reconfigurable metasurface apertures have been demonstrated at X (8-12 GHz)[14] and K (18-26.5 GHz)[15] bands, and are very likely achievable at W (75-110 GHz) band in the near term. WPT systems operating within any of these bands are within the realm of possibility depending on the particular application. Power harvesting elements at these wavelengths, such as rectennas, have been demonstrated, but high conversion efficiency circuits may require additional development. Assuming optimal conversion efficiency, we can obtain estimates of the useful range of a Fresnel zone WPT system as a function of aperture size and frequency of operation, based solely on the ideal field patterns expected to fill the aperture.

The actual metasurface antenna implementation of a dynamic aperture will have limitations that arise due to the finite size of the metamaterial elements (leading to a subwavelength sampling of the aperture), as well as their inherent dispersive characteristics. Using a well-established model for these elements that describes both their dispersive and radiative properties, it is possible to determine the actual focal spot size and shape, including aberrations introduced by any phase or amplitude limitations inherent to the elements. Phase or amplitude patterns on the metasurface aperture that steer the focal spot throughout the volume of coverage can be determined using holographic techniques, so the effective power transfer efficiency can be studied as a function of the receiver location and orientation.

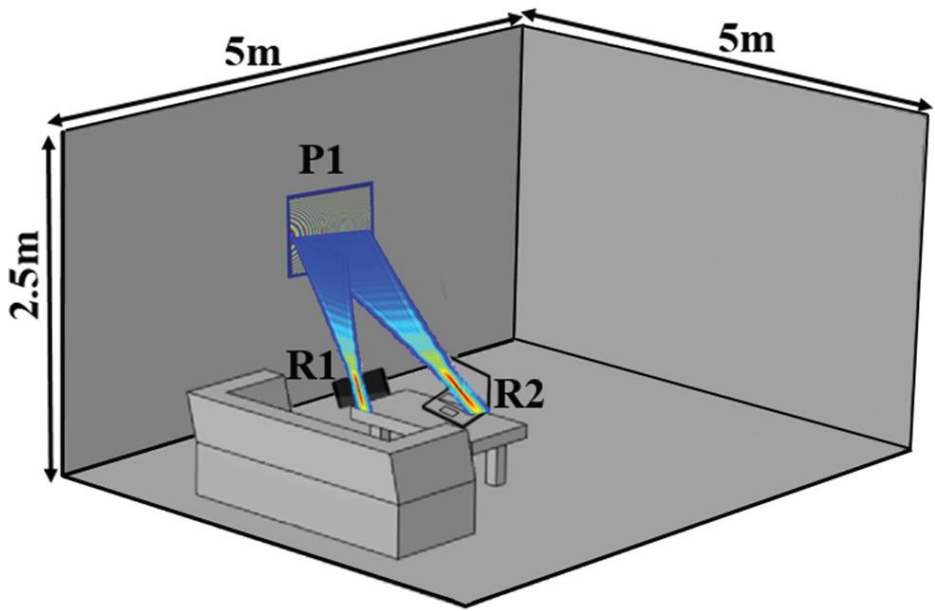

FIG. 1. Potential usage scenario for a Fresnel zone wireless power transfer system.

There are many potential usage opportunities for WPT schemes operating in the near-zone. Since the location of Fresnel zone (and realizable focus spot size) depends only on the size of the transmit aperture and the wavelength of operation, many different scenarios can be considered. One possible scenario is presented in Fig. 1, which shows an aperture being used to beam power wirelessly to electronic devices within the confines of a room. The advantage of such a scheme is that devices—such as cell phones, laptops, computer peripherals, gaming controllers or consoles, watches, radios, small appliances—can be positioned anywhere within the line of sight of the source to receive power, requiring no cables or charging stations. For such a scheme to function, each device must be discoverable and locatable by some separate wireless system, which could be built into the protocols of a near-zone system. That is, each device must signal its presence in the room, and communicate its location and orientation with respect to the transmit aperture. In addition, if the device is being powered, it must indicate to the transmit aperture that power is being delivered, such that the system shuts down if the direct path is blocked for any reason (e.g., if a person passes between the aperture and device). This safety interlock system would be essential to accommodate human safety requirements. Since our goal here is to consider the general viability of near-zone power transfer from a power transfer efficiency perspective, we do not consider further the issues of safety and other related protocols that would ultimately become a system engineering topic.

For the purposes of the analysis presented here, we consider powering devices within a room of dimensions of 5 m x 5 m x 2.5 m, requiring an aperture large enough that all points within the room are also within the Fresnel region of the transmit aperture. As shown in Fig. 1, a single transmit aperture (P1) focuses RF power to several electronic devices within the room, such as R1 and R2. The devices are placed at arbitrary locations in the room, at different focal depths with respect to the transmit aperture, as would be expected in real use scenarios. The transmit aperture must thus be capable of powering targets at different depths and angles, as well as potentially powering multiple receivers from a single aperture. This functionality implies a dynamic aperture capable of creating a tight focus and adjustable focal length.

## II. POWER TRANSFER IN THE FRESNEL ZONE

The efficiency of a WPT system in either the Fresnel ($< 2D^2/\lambda_0$) or the Fraunhofer ($> 2D^2/\lambda_0$) regime depends predominantly on the effective aperture sizes of the source and receiver, as well as the free space wavelength. Given our choice of Fresnel region operation the far-field propagation model is not valid. Instead, in the Fresnel region, the aperture behaves like a lens, able to concentrate the transmitted energy to a volume defined by its point spread function (PSF). As a starting point, we assume the aperture behaves as a Fresnel lens, and apply Gaussian optics to characterize the expected field patterns. We consider the line of sight case where a source beam will travel without encountering obstructions. The spatial electric field distribution corresponding to a Gaussian beam can be described analytically using the expression[16,17]

$$\frac{E(x,y,z)}{E_0} = \frac{w_0}{w} e^{-\frac{r^2}{w^2}} e^{-i\frac{k_0 r^2}{2R}} e^{-i(k_0 z - \phi)}, \tag{1}$$

where $r = \sqrt{x^2 + y^2}$ is the distance from the center of the focus. In this expression $w$ is the beam waist, given by

$$w(z) = w_0 \sqrt{1 + \left(\frac{z}{z_r}\right)^2}, \tag{2}$$

In Eq. 2, $w_0 = w(0)$ is the beam waist at the focus, and $z_R$ is the Rayleigh length, defined as

$$z_R = \frac{\pi w_0^2}{\lambda_0}, \tag{3}$$

Finally, $\phi$ is the Guoy phase shift and $R$ is the radius of curvature, which have expressions

$$\phi(z) = \tan^{-1}\left(\frac{z}{z_R}\right)$$
$$R(z) = z\left(1 + \frac{z_R^2}{z^2}\right), \tag{4}$$

Assuming that the fields are focused from a lens of diameter $D$ and focal length $z_0$, the minimum beam waist can be calculated as

$$w_0 = \frac{4}{\pi} \frac{z_0 \lambda_0}{D \cos^2(\theta)}. \tag{5}$$

where $\theta$ is the angle between the optical axis of the aperture and the vector defined from the aperture center to the focal point. We make an initial approximation that the beam waist increases as the square of the cosine of $\theta$, accounting for the change in effective focal length and decreased aperture size off-axis (on-axis corresponds to $\theta = 0°$). Eq. 5 is derived in the appendix.

Eq. 1 shows that the fields of a focused beam tend to be tightly confined laterally around the focus, but extend along the propagation direction by a distance corresponding to the Rayleigh length. Thus, for short focal lengths relative to the aperture dimension, the fields tend to be confined in all dimensions; however, for larger focal lengths, the Rayleigh length tends to be larger and the fields spread out along the propagation direction. Plots of the intensity of the focused beam for several values of $z_0$ are shown in Fig. 2. For these illustrative plots, the aperture size is chosen as $D = 1$ m, and the frequency as $f = 10$ GHz.

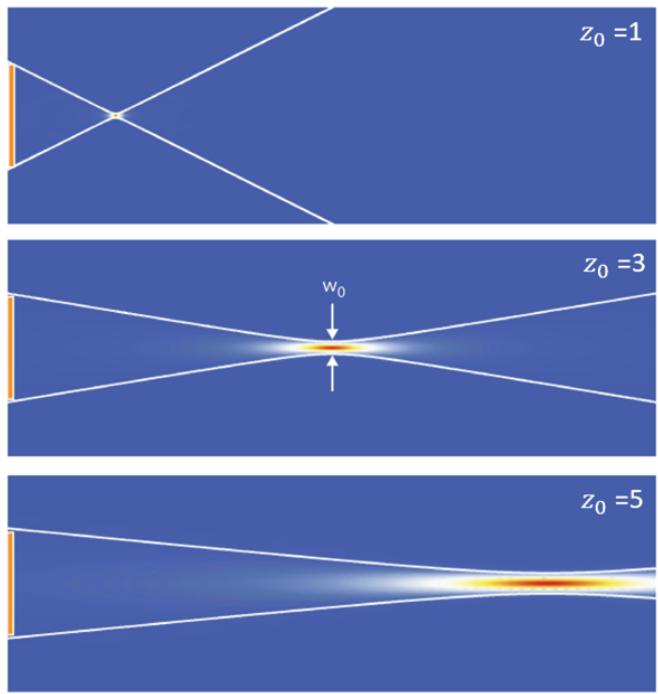

FIG. 2. Intensity plots of a focused beam from an aperture of dimension $D=1$ m, at a frequency (f) of 10 GHz, plotted for several values of focal length ($z_0$). Solid white curves are plots of the beam waist w(z).

Since the minimum beam waist increases with focal length, a straightforward design consideration for the system is that the transfer efficiency must be optimized at the farthest desired point from the aperture. For this study, we assume the smallest receive aperture to have dimension of $D_{Rx} = 3.0$ cm, so that a simple measure of transfer efficiency can be taken as $\eta = D_{Rx}/w_0$. This simplistic model likely represents a best-case limit of efficiency.

Using Eq. 5, we can perform an initial study of the beam waist versus aperture dimension and frequency as a means of assessing the initial constraints of a WPT system. Such a study is presented in Table I, which provides the minimum beam waist at a focal length of 5 m for several values of aperture size and operating frequency.

Inspection of Table I shows that, as expected, larger aperture and higher frequency provide the smallest beam waists. Given the receive aperture size considered here ($D_{Rx} = 3.0$ cm), there are a variety of combinations of transmit aperture size and frequency that should optimize efficiency. Assuming a transmit aperture footprint of no larger than 1 m$^2$, Table I shows that frequencies of 60 GHz or higher should provide reasonable transfer efficiency over the volume considered. The eventual choice of frequency will likely be determined by the availability, cost and power conversion efficiency of the RF power source and the rectifier or energy harvester at the receiver. At present, for example, low cost solid-state sources are emerging into the market for bands at 60 GHz and 77 GHz, due to the demand in automotive radar and other large market applications.

TABLE I: Beam waist (in cm) as a function of frequency and aperture size, using Eq. 1. The shaded regions might be considered acceptable for this study.

|  |  | D(m) | | | |
|---|---|---|---|---|---|
|  |  | 1 | 2 | 3 | 4 |
| f (GHz) | 20 | 9.54 | 4.78 | 3.18 | 2.38 |
|  | 40 | 4.78 | 2.38 | 1.60 | 1.20 |
|  | 60 | 3.18 | 1.60 | 1.06 | 0.80 |
|  | 80 | 2.38 | 1.20 | 0.80 | 0.60 |
|  | 100 | 1.90 | 0.96 | 0.62 | 0.48 |

The possibility of achieving even smaller beam waists at much higher frequencies, such as THz, infrared or even visible, can also be considered. However, highly efficient rectennas (rectifying antennas) can be challenging to design at these frequencies[18]; sources are expensive and not readily available; and the power density in such highly collimated systems can exceed human safety limits.

The beam waist and Rayleigh length are the critical parameters for the description of a Gaussian beam, and can be used to generate general scaling arguments for various quantities of interest. For example, assuming all power incident on the receive aperture is recovered, the relative size of the receive aperture to the beam waist should be the only relevant quantity in terms of describing efficiency. Fig. 3a shows how efficiency scales with the ratio of $\eta = D_{Rx}/w_0$. As expected, the transfer efficiency reaches a relatively high value (>80%) when this ratio is unity, and approaches 100% as the aperture area increases beyond unity. This curve is invariant with respect to focal length, frequency or other parameters. Likewise, the transfer efficiency drops off away from the focus as a function of distance along the propagation axis, as shown in Fig. 3b. If the efficiency is plotted against the unitless parameter $z/z_R$, then a universal curve results.

A couple of items should be noted here. We have adopted a fairly simple definition of transfer efficiency that will be applied throughout this analysis. It is relevant, however, to consider in a little more detail what might be the upper limit on possible free-space power transfer efficiency within the Gaussian beam approximation, which should be agnostic to the manner in which the beam is created and absorbed.

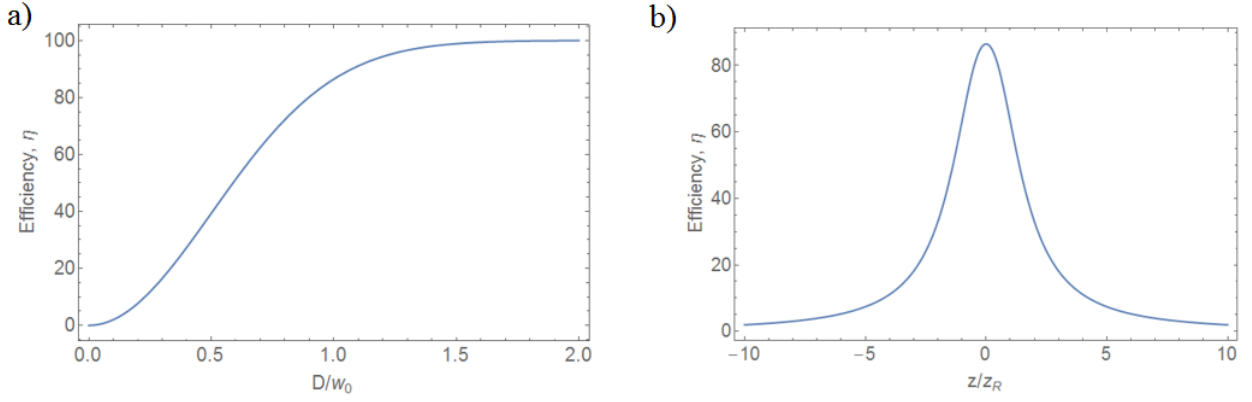

FIG. 3. (a) Transfer efficiency assuming the receive aperture is placed at the focus, as a function of aperture size (scaled to the beam waist). (b) Transfer efficiency assuming a receive aperture of size $w_0$, where ($D/w_0 = 1$), as a function of scaled distance away from the focus.

The upper bound estimate on transfer efficiency is based on the notion that a Gaussian beam is a free-space mode with a known power flux distribution, described entirely by its minimum waist size $w_0$ and the focal point. The transmitter and receiver apertures couple energy in and out of this mode. For simplicity, consider circular apertures of radii $R_{Tx, Rx}$, both concentric with the beam axis and perpendicular to it. Based on the energy density distribution of the beam (found from Eq. 1), and the notion that the energy is localized with spatial accuracy of $\sim \lambda_0$ (which becomes a negligibly small scale in the large aperture limit), the coupling efficiency between an aperture and the beam cannot exceed

$$\eta_{Tx,Rx} = 1 - e^{\frac{-2R_{Tx,Rx}^2}{w^2(z_{Tx,Rx})}}. \tag{6}$$

where $w(z_{Tx,Rx})$ is the beam waist radius in the plane of Tx or Rx apertures, respectively, as given by the beam waist equation (Eq. 2). Then, the upper limit on energy transfer efficiency between these apertures is given by

$$\eta = \eta_{Tx}\eta_{Rx} \tag{7}$$

From a practical perspective, the dimensions such as aperture diameters and the distance between them are given, and beam parameters such as $w_0$ and center coordinate (relative to either aperture) can be optimized to maximize the efficiency (Eq. 7). It can be easily shown that, as a function of the focal point position, the optimum of (Eq. 7) is always achieved when the focal point is co-located with the smaller of the two apertures, typically the receiver. However, the relationship between the optimum waist and the receiver size is less trivial. Fig. 4a shows that efficiency limit (Eq. 7) usually has a maximum as a function of $w_0$ (for a fixed $R_{Rx}$); this is a consequence of a trade-off between decreasing efficiency of beam absorption on the receiver and increasing efficiency of its creation as the minimum waist size goes up (wider-waist beams are easier to create). The optimum value of $w_0$ is given by an algebraic equation, whose numerical solution is plotted in Fig. 4b. One can see that, although optimal beam waist radius is always of the order of the receiver radius, the exact value of best $w_0/R_{Rx}$ is not unity, and it depends on other geometric parameters, such as the transfer distance and the receiver radius itself.

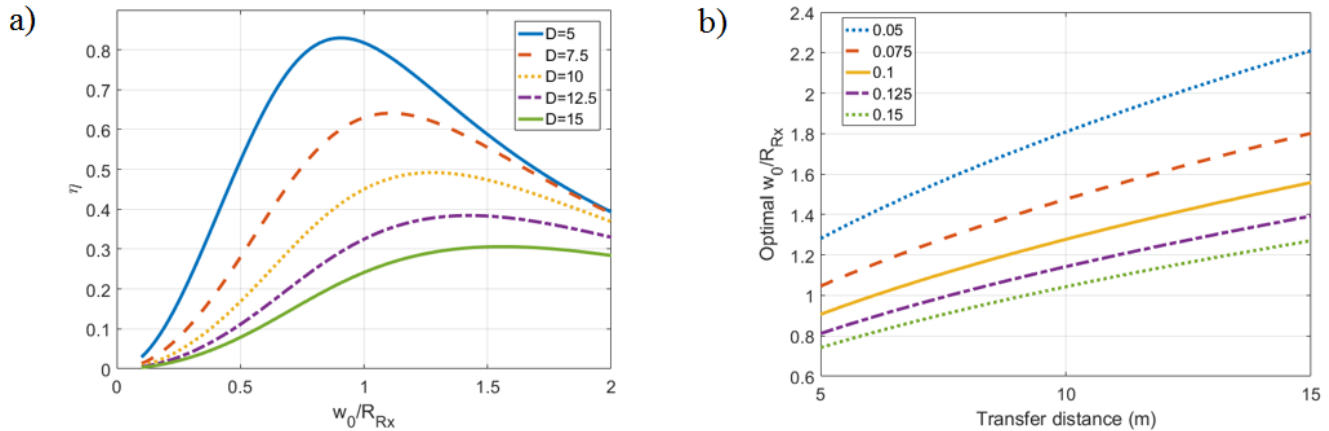

FIG. 4. (a) Maximum transfer efficiency as a function of minimum beam waist for several different transfer distances (D, in meters). (b) Optimized minimum beam waist vs transfer distance, for several different Rx radii (in meters). Tx radius 1 m and frequency 5.8 GHz are assumed. Minimum beam waist is normalized to Rx radius.

In what follows, we find excellent qualitative and quantitative agreement with the Gaussian optical approximation, which supposes a focal spot with fields that decay exponentially away from the spot center. In reality, the effects of diffraction and finite sampling lead to the appearance of side lobes that surround the main beam. A cross-range plot over the focal spot reveals additional field oscillations that are outside of the Gaussian model. While the width of the main spot is well-predicted by Eq. 5, additional lobes emerge that will contain some small amount of power. The side lobes are a simple consequence of filling the aperture more-or-less uniformly with a constant field (modulated by the holographic patterning), and are independent of the aperture size. We consider a simple case of on-axis focusing and present the 1-D cross-range plots of the normalized electric field intensity at the focal plane. For this study, aperture sizes of 0.5 m, 1 m and 2 m are considered. As shown in Fig. 5, the width of the main lobe and the side lobe both change with the change in the aperture size (the plot shown in Fig. 5 is zoomed in towards the center to focus more on the behavior of the main lobe and the side lobes). In Fig. 5, we also analyze the side lobe level as a function of aperture size. The ratio between the first side lobe level to the main lobe is calculated to be around -14 dB shown by the black dotted line in Fig. 5, roughly in accordance with simple diffraction theory for a uniformly illuminated aperture. The sampling on the focal plane and the aperture plane considered in this analysis is $\lambda_0/2$. The trade-off between side lobe levels and main peak width are well-known[8]. The use of tapers and other methods of optimization could be applied to lower the side lobes, but already the side lobes can be expected to have minimal impact on the power transfer. In the simulations presented below, the side lobes are present, and their effects are included in the efficiency calculations.

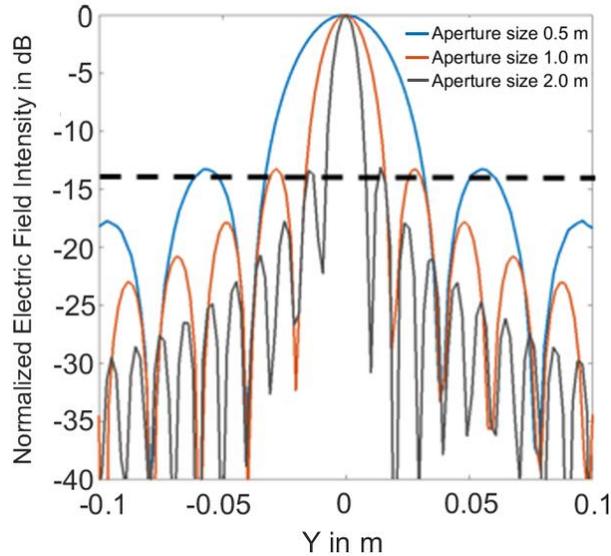

FIG. 5. 1-D cross-section plot (zoomed version) of the normalized electric field intensity at the focal plane for on-axis focusing for different aperture dimensions.

**III. PRACTICAL IMPLEMENTATIONS**

**A. Aperture architecture**

In any feasible implementation of a near-zone WPT scheme, the source aperture is likely to take the form of a flat panel that can be wall- or ceiling-mounted and fairly unobtrusive. Wall or ceiling mounting offers the important advantage of line-of-sight propagation to most points in a room. For conventional lenses, even in the most favorable of circumstances, some level of aberrations would be introduced into the beam due to the inherent limitations of planar optics. While the characteristics of the dynamically reconfigurable sources considered here are different from static lenses, additional imperfections in spot size and other metrics can be expected, since the aperture will necessarily be sampled discretely with components that may have limitations in their phase or amplitude control range.

One means of forming a dynamic aperture is a phased array or electronically scanned antenna[19,20]. A phased array consists of an array of radiating elements, each element containing a phase shifter and possibly an amplifier. The radiating elements are positioned at distances of roughly half the free space wavelength apart. If full control over phase and amplitude is available, then it follows from Fourier optics that an electronically scanned antenna has the capability to produce any far-field pattern. However, from the standpoint of a high efficiency WPT system, phased arrays or electronically scanned antennas are not an optimal solution since each of the radiating modules requires external bias power (beyond the wireless power to be transmitted). Power consumption in array control systems can be substantial, easily exceeding the power being transferred to small devices.

A second issue associated with phased arrays is that sampling the aperture at Nyquist limits (roughly half the free space wavelength) can result in arbitrary far-field patterns, but not arbitrary patterns in the near-zone. Thus, it is necessary to subsample the aperture relative to Nyquist, requiring far more active elements to achieve dynamic focusing.

An alternative architecture for dynamic focusing is that of the metasurface aperture, which is—in contrast to electronically scanned antennas and phased arrays—a largely passive device that achieves reconfigurability via dynamic tuning of metamaterial resonators. The details of the metasurface architecture are beyond the scope of the present study, but we assume some of the metasurface aperture constraints in the consideration of more realistic implementations.

**B. Holographic aperture design**

As a next level of approximation, we consider the formation of focal spots using an aperture over which any field distribution can be obtained. In this section, we assume that the aperture can be sampled as finely as desired, so that the limitations associated with a flat and finite aperture are explored. To determine the fields everywhere in the region of interest, the fields at the aperture can be propagated using the angular spectrum method (ASM)[16]. In this method, a Fourier transform is taken of the fields on the aperture, resulting in a set of coefficients corresponding to an expansion in plane waves. Each of these plane wave components is then propagated a given distance along the propagation direction, where an inverse Fourier transform can be taken to find the field distribution over the plane.

In this work, we assume that an arbitrary field distribution (both amplitude and phase) can be created over the aperture plane, to varying approximations, that will produce a focused spot. We determine the required field distributions by designing a hologram[21-22]—the recorded interference pattern between a reference beam and the scattered complex fields from an object located within the Fresnel region. To arrive at the required amplitude and phase distribution of the aperture field, we construct the aperture field by interfering a point source with a uniform plane wave. Taking the center of the aperture as the origin of the coordinate system, a point source located at the position $(x_0, y_0, z_0)$ will produce the aperture field

$$E(x,y,0) \propto \frac{e^{ik\sqrt{(x-x_0)^2+(y-y_0)^2+z_0^2}}}{\sqrt{(x-x_0)^2+(y-y_0)^2+z_0^2}} \quad . \tag{8}$$

Eq. 8 provides the amplitude and phase distribution needed to design a holographic pattern that will produce a point source of diffraction-limited extent. This initial field distribution can then be back-propagated to determine the fields everywhere in the region of interest. As a practical matter, we find that the amplitude variation is not of great importance in reproducing the point source, so we use only the phase distribution in the following analysis. Examples of the phase hologram are shown in Fig. 6, as well as the focused spots produced by these apertures. The hologram fringes must be faithfully reproduced to minimize aberrations, which means we must sample the aperture so as to capture the spatial variation. For the simulations in this section, a sampling of $\lambda_0/8$ was used.

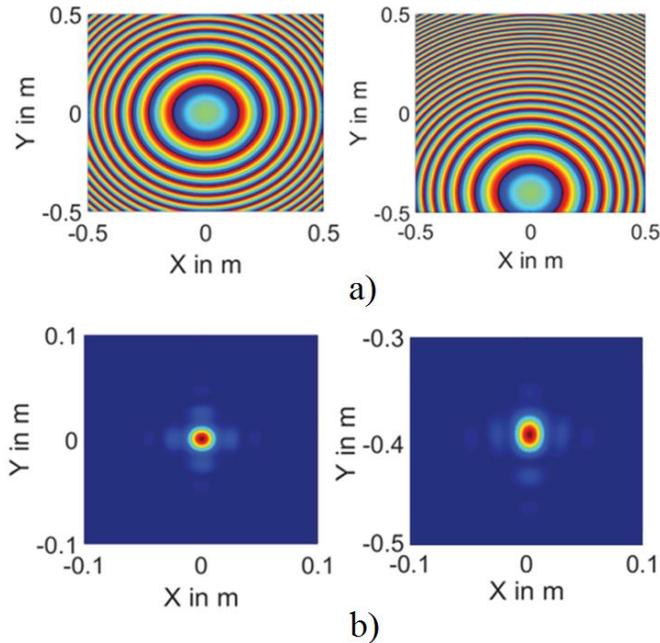

FIG. 6. Cross range 2D plots (a) Phase distribution for both on- and off-axis case at the aperture plane(b) Focused spots (zoomed in) at focal plane for both on- and off-axis focusing.

Initially, we investigate focusing with an ideal holographic transmit aperture and a receive aperture placed some distance away. Fig. 6 (a) shows the phase distribution required on the aperture plane (on-axis and off-axis), with a 2D intensity plot of the focused spots produced by these apertures. For both simulations, $z_0=5$ m. The offset value chosen for off-axis focusing is for illustrative purposes only. Comparing the on-axis and off-axis focused intensity patterns, it can be seen that the spot for the off-axis elongates in the cross range direction; that is, the beam waist for the off-axis case slightly increases when compared to the on-axis focusing.

The increase in beam waist for off-axis beams is expected due to aperture loss. To confirm the behavior, we compute the beam waist at the focus for an ideal holographic aperture of dimension $D_{Tx}=1$ m. The transmit aperture is designed to produce a focus at a distance of $z_0=5$ m. Since the selection of the frequency band mainly depends on commercially-available sources, an operating frequency of 77 GHz (corresponding to the automotive radar band) is considered for the analysis presented. The beam waist at the focal plane is taken as the diameter at which the intensity has decreased to $1/e^2$ or 13.5% of its peak value.

The beam waist as a function of off-axis angle (offset along y-axis) is shown in Fig. 7a for both the simulations and the approximate analytical Gaussian formula of Eq. 5. The coordinates of the focal points are selected such that the total distance of the focal point to the center of the aperture is constant ($r = \sqrt{x_0^2 + y_0^2 + z_0^2}$).

Using a hypothetical receive aperture of dimension $D_{RX} = 3$ cm, we can assess power transfer efficiency for off-axis beams. As can be seen in Fig. 7b, the transfer efficiency decreases as the beam offset angle $\theta$ increases, as expected due to spillover losses.

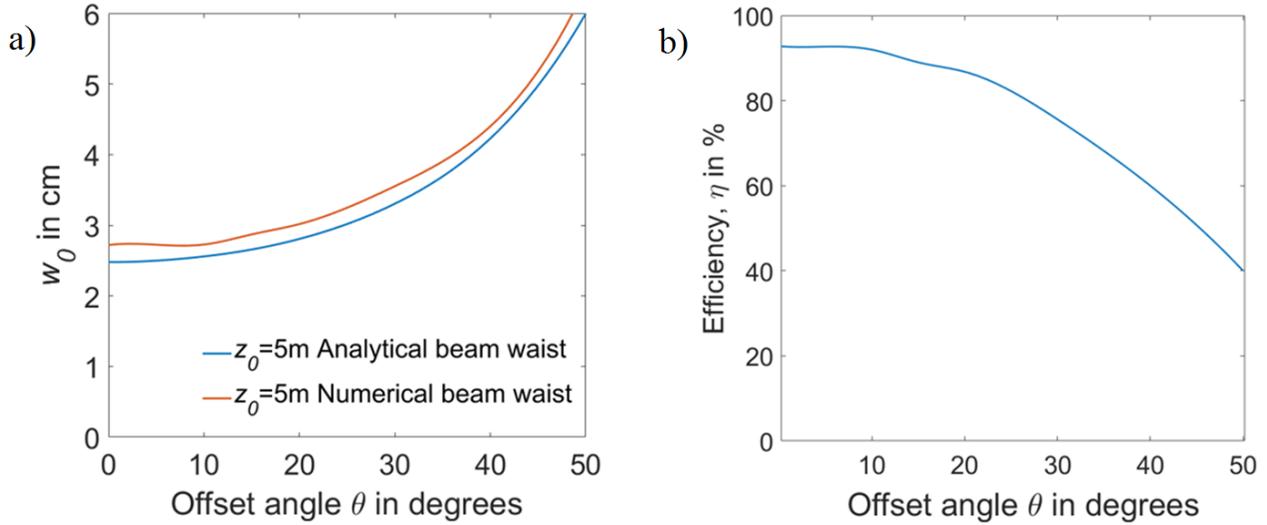

FIG. 7. (a) Analytical and numerical beam waist as a function of offset angle θ for a focal length $z_0 = 5$m . (b) Power transfer efficiency as a function of offset angle ($D_{Rx} = 3$ cm).

If one assumes a fixed receive aperture of minimum dimension (assumed to be $D_{Rx} = 3$ cm), then Eq. 5, as well as the numerical results, suggest a coverage area map can be formed that includes all regions of interest for which the beam waist will be small enough for a desired level of power transfer efficiency. An analytical estimate of this coverage zone can be extracted from Eq. 5 by examining contours in the y-z plane of constant beam waist. For a constant beam waist, Eq. 5 leads to the following:

$$y^2 + \left(z - \frac{d_c}{2}\right)^2 = \left(\frac{d_c}{2}\right)^2. \tag{9}$$

where the constant $d_c$ is

$$d_c = D \frac{\pi w_0}{4 \lambda_0}. \tag{10}$$

Eq. 9 shows that contour of constant beam waist is a circle in the y-z plane centered at $z = d_c/2$ and with diameter $d_c$. Thus, for a given aperture size and operating wavelength, a desired coverage range ($d_c$) can be selected and Eq. 8 used to determine the beam waist needed. In the example we study here, the relevant parameters $D = 1$ m, $\lambda_0 = 4$ mm and $w_0 = 3$ cm. suggest a coverage diameter of 6 m. All beam waists within a given contour are smaller than this value, so that the least power transfer efficiency occurs at the periphery of the coverage region. Plotting Eq. 5 results in a coverage map such as that shown in Fig. 8a.

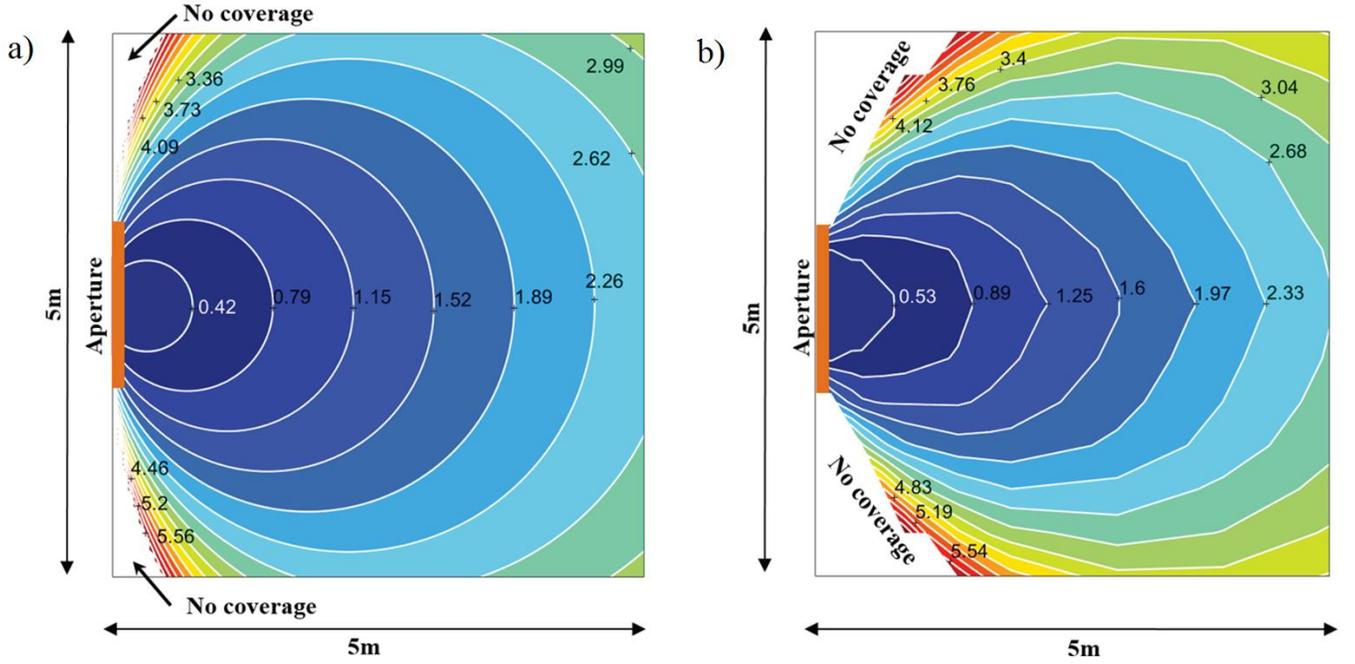

FIG. 8. Illustrative coverage plot of beam waist as a function of off axis angle and focal distance in a room (5m x 5m, top view) (a) Analytical equation (b) Numerical calculations using simulations.

Since we expect the angular behavior predicted by Eq. 5 to be very approximate, we also directly compute the coverage map using the numerical method described above. From this map (Fig. 8b) we see that the contours of constant beam waist are somewhat more elliptical, yet nevertheless fairly close to the simple coverage map predicted by Eq. 9.

**C. Phase and Amplitude Constrained Holograms**

For the studies conducted above, we assumed that an arbitrary field distribution (in both amplitude and phase) could be imposed across the aperture, resulting in diffraction-limited focal spots being generated. Although point-by-point simultaneous control of the phase and amplitude over an aperture can be achieved to a considerable extent in active electronically scanned antennas, such systems are not yet economically viable for larger-volume, low-cost applications such as the WPT scenario envisaged here. The alternative dynamic metamaterial aperture provides a low-cost, manufacturable alternative platform, but comes with certain limitations. In particular, it is not possible to control the phase and amplitude of a resonator-based metasurface aperture independently. The resonance of the metamaterial resonator element possesses a Lorentzian relationship between the phase and amplitude, offering a constrained control space. Moreover, for a single resonator, the maximum range of the phase is between -90° and +90°, placing an immediate limitation on the field distribution over the aperture[23]. In practice, because the amplitude of a Lorentzian falls off substantially away from the 0° point, the useful phase range is likely substantially smaller. The possibility exists of combining both an electric and a magnetic resonator into the same radiating elements, which would allow the full 360° phase range to be accessed; however, these Huygens surfaces are considerably more difficult to design and may be more subject to resistive losses and unwanted inter-element coupling. We will consider Lorentzian constrained holograms later.

In this section, we first investigate the potential performance of the holographic aperture for WPT in the presence of phase limitations. Because the dependence of the image produced by a hologram is typically only weakly dependent on the magnitude, we first consider either amplitude-only or phase-only holograms to assess focusing capability.

We form the desired hologram in the same manner as in Section III.B above, conceptually interfering a plane wave with the spherical wave from a point source at $z_0$ over the aperture plane, resulting in the specification of the required complex field distribution over the aperture. An amplitude hologram can be realized by enforcing a binary mask over the aperture, achieved by treating each sampling point as either transparent or opaque; that is, each point on the aperture controls the amplitude in a binary fashion, introducing no phase shift. The result of the binary amplitude hologram is shown in Fig. 9a, where the hologram has been designed to produce a focal spot at an angle of 15° from the normal to the aperture at a distance of $z_0 =$ 5m in range. Not surprisingly, the aperture produces a zeroth-order diffracted beam and several other diffracted beams, such that energy is lost from the main focus and additional hotspots are created in the region of interest. The locations of these additional beams and associated focal points can be determined using simple diffraction theory, and agree with the patterns found from the simulation. The scenario illustrated in Fig. 9 can be considered a worst-case scenario, since no attempt was made to optimize the aperture distribution. The amplitude mask was created from the ideal hologram by setting points within a particular phase range to transparent, and points outside that phase range to opaque. Two simple binary amplitude designs were considered here: In Aperture A, those regions with phase between 0° and 90° were set to transparent (all other regions opaque), while in Aperture B those regions with phase between 0° and 45° were set to transparent (all other regions opaque).

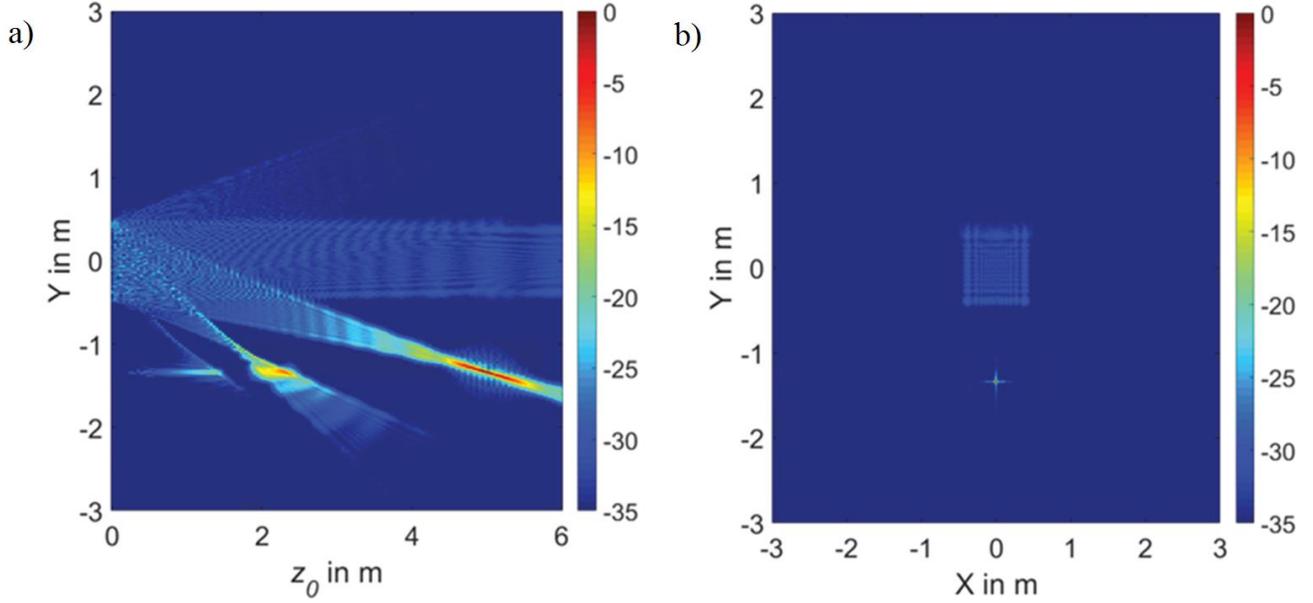

FIG. 9. Field distributions for an amplitude holographic lens designed to produce a focus at 15° away from the normal. (a) Intensity distribution in dB of the y-z plane showing higher diffraction orders (b) Intensity distribution in dB of the x-y plane (cross range) plot of off-axis focusing with the zeroth order mode at the center.

The simulations were conducted for an aperture plane of size $L=6$ m containing a transmit aperture of size $D_{Tx}=1$ m, as shown in Fig. 9a. The focus was chosen to be at $z_0 = 5$ m, at an angle of 15° from the aperture normal. The fields are determined over the transmit aperture in the manner described above, and set to zero elsewhere on the plane of the overall aperture. For the two apertures considered, A and B, the overall transfer efficiency was determined, as summarized in Table II. As above, in this case one contributor to transfer efficiency is the ratio of the intended receive aperture ($D_{Rx}$) to that of the beam waist at focus, or $\eta$. In addition, we define the ratio, $\alpha$ as the total power radiated by the aperture to the total power in the first order (desired) focused beam. Thus, the overall efficiency of power transfer to the device is $\eta/\alpha$. The total power is calculated by integrating the Poynting vector over the entire surface of the domain (at $z_0 = 5$ m). A value of $\alpha =1$ indicates no presence of higher order modes.

TABLE II: Overall efficiency of amplitude-only holograms for off-axis focusing and ratio $\alpha$ for different binary holograms.

| Aperture | Overall efficiency (%) | $\alpha$ |
|---|---|---|
| A: 0° to 90° | 14.4 | 2.4 |
| B: 0° to 45° | 8.9 | 2.8 |

As shown in Table II, $\alpha$ is above 1 for both cases considered, and the efficiency is generally low, indicating that significant power is lost to higher order diffracted modes. With additional optimization, it is certainly possible to suppress some or all of the diffractive orders within an amplitude-only hologram, especially if the amplitude is allowed to take a range of values (grayscale) rather than just binary[24,25]. We do not consider such optimization here.

A phase hologram can be realized by starting from the ideal hologram specification, obtained from the interference between a plane wave and a point source at $\alpha$. We assume the phase at each sampled point on the aperture can be controlled. An ideal phase hologram would allow the phase to vary from -180° to +180° (aperture C), leading to essentially perfect efficiency, as shown in Table III.

For a phase only hologram, any limitation of the phase range to less than 360° will result in an imperfect hologram and degraded focusing performance; moreover, the inevitable phase discontinuities that result can produce scattering into the higher order diffraction modes. To analyze the effect of limiting the phase, we consider again the case of off-axis focusing (an angle of 15° from the aperture normal), limiting the phase values across the aperture to lie within a restricted range, as summarized in Table III. The simulation domain for these examples is identical to that used for the amplitude-only hologram. In these simulations, where the phase of the ideal aperture is required to be smaller than the lower limit of the indicated phase range, the phase was set equal to the lower limit. Where the phase of the ideal aperture is required to be larger than the upper limit shown, the phase was instead set equal to the upper limit. As Table III shows, constraining the phase range reduces the overall efficiency.

The field patterns for phase-only holograms with various phase constraints are shown in Fig. 10. Again, the aperture is designed to focus at $z_0 = 5$ m at an off-axis angle of 15° from the normal to the aperture. Due to the phase discontinuity introduced on to the aperture, other diffraction orders occur, resulting in loss of power from the main order or focus. The scenarios considered in Fig. 10 correspond to the selected apertures summarized in Table III (except for Aperture C). With the full 360° of phase values, no higher diffraction orders were observed.

TABLE III: Overall efficiency of phase-only holograms for off-axis focusing and ratio $\alpha$ for different phase limited apertures.

| Aperture | Overall efficiency (%) | $\alpha$ |
|---|---|---|
| C: -180° to 180° | 78.1 | 1.00 |
| D: -135° to 135° | 74.3 | 1.02 |
| E: -90° to 90° | 52.4 | 1.38 |
| F: -60° to 60° | 28.9 | 2.40 |
| G: -30° to 30° | 8.30 | 7.60 |

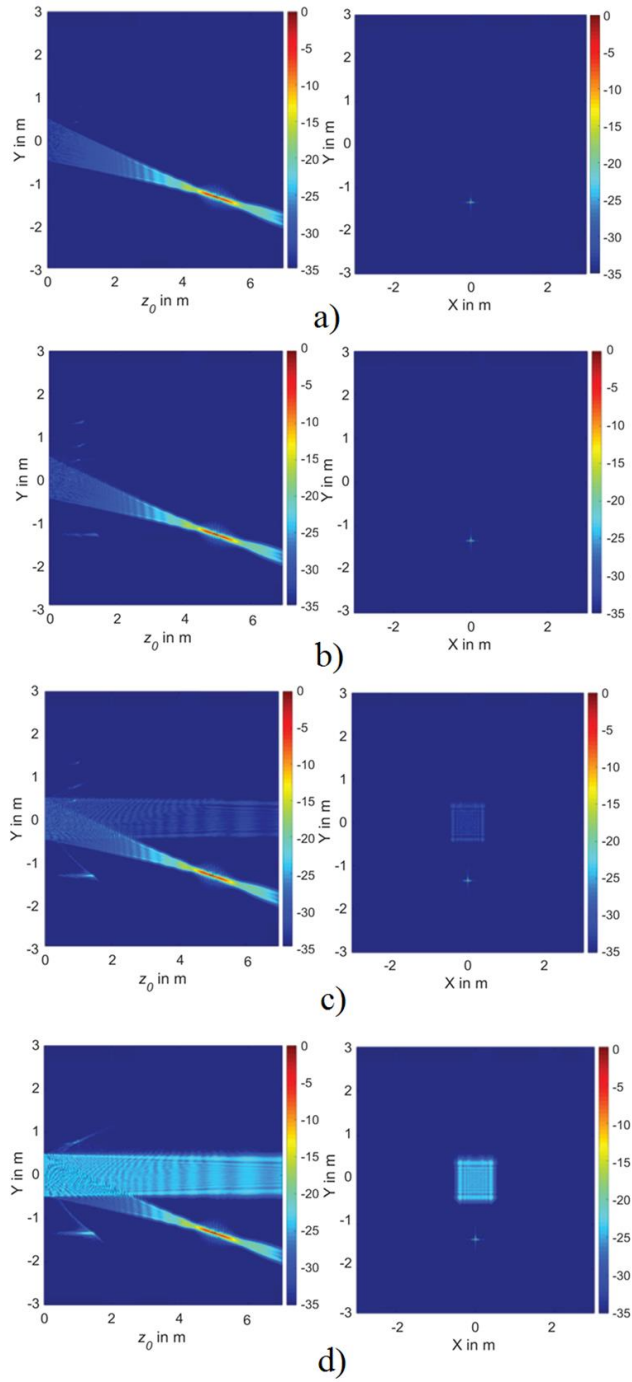

FIG. 10. Intensity distributions (dB) in the y-z plane (range) and x-y plane (cross-range) for a phase holographic lens designed to produce a focus at 15° away from the normal. (a) Aperture D (b) Aperture E (c) Aperture F (d) Aperture G.

The beam waist of the main beam remains relatively constant for the various phase holograms, so that the drop in efficiency can be associated with the power loss in the other diffracted orders. The increasing value of $\alpha$ in Table 3 indicates this loss. The ideal phase-only hologram produces no higher diffraction orders, and thus has a value of $\alpha = 1$. In aperture D, phase values between -135° to +135° are allowed, with the remaining sampling points, or pixels, set to either the upper or lower bounds of the phase limits considered. Despite aperture D having a significantly restricted phase range, the overall efficiency remains high and higher order modes are not important. Similarly, in aperture E, phase values between -90° to +90° are allowed. The value of $\alpha = 1.38$ corresponds to the increased presence of unwanted diffracted beams as expected. Apertures F and G show a significant increase in α because of the increase in the power lost to the other orders, mainly the zeroth order (DC component), as shown in Fig. 10c and Fig. 10d respectively. The zeroth order becomes stronger as the phase range on the aperture is reduced. Aperture G corresponds to the worst case scenario of the phase limitations considered, resulting in a largest value of α. The holograms shown here are meant to be illustrative, and have not been optimized. Diffraction into unwanted orders can potentially be removed or at least suppressed through various optimization techniques[26] such as iterative algorithms like Gerchberg-Saxton or Hybrid Input-Output[27].

### D. Lorentzian Constrained Holograms

We have so far considered the analysis of amplitude-only and phase-only holograms, but if we consider a practical metasurface antenna, the amplitude will be linked to the phase through the dispersion relation of the Lorentzian resonance. The metasurface antenna can be modeled as a collection of polarizable magnetic dipoles[28], with each dipole possessing a polarizability of the form

$$\alpha_m = \frac{iF\omega^2}{\omega^2 - \omega_0^2 + i\gamma\omega} \ . \tag{11}$$

From Eq. 11, the phase of a metasurface element is related to its resonance frequency according to

$$\tan(\theta) = \frac{\omega^2 - \omega_0^2}{\omega\gamma} \ . \tag{12}$$

where the $\gamma$ is a loss term, $\omega_0$ is the angular resonant frequency, and $F$ is a coupling factor, which we can set equal to unity for the discussion presented here.

To form a desired phase distribution using metasurface elements, the resonance frequency of the element can be controlled by some method. However, by tuning the resonance frequency, the amplitude of the polarizability is also determined having the form

$$|\alpha_m| = \frac{F\omega^2}{\sqrt{\left(\omega^2 - \omega_0^2\right)^2 + \gamma^2\omega^2}} \ . \tag{13}$$

For a desired phase pattern $\theta$ over the aperture required to create a focus, determined by modifying the resonance frequency at each point, an amplitude pattern will necessarily be imposed defined on Eq. 12. Inserting Eq. 12 into Eq. 13 yields a relatively simple expression linking the phase and amplitude at each point:

$$|\alpha_m| = \frac{F\omega}{\gamma}|\cos\theta| . \qquad (14)$$

Eq. 14 shows that the amplitude is proportional to the absolute value of the cosine of the phase (with 0° occurring at resonance), falling to zero at the extreme values (-90° or +90°).

To investigate the impact of the Lorentzian constrained aperture, we again consider the same scenario as above, with the aperture designed to focus at $z_0 = 5$ m at an off-axis focusing (an angle of 15° from the aperture normal). Here we consider holograms formed by limiting the phase values across the aperture to lie within a restricted range, as summarized in Table IV, while the amplitude at each point is determined from Eq. 14. The simulation domain for these examples is identical to that used for the amplitude-only and phase-only holograms. As shown in Table IV, limiting the phase reduces the overall efficiency and produces other diffraction orders. The scenario considered in Fig. 11 corresponds to Aperture H as shown in Table IV. The beam waist for the apertures considered in Table IV are constant and the loss in overall efficiency is due to the power lost from the first order mode to the other diffracted orders. Aperture H can be considered as the best case for a metasurface antenna (without further optimization), since it extends the full phase range of -90° to 90°. The overall efficiency is around 34.2 %. We then consider aperture I, in which the phase range is further restricted, resulting in further reduction in the overall efficiency due to the increase in the α term as shown in Table IV.

Table IV: Overall efficiency of a metasurface for off-axis focusing and $\alpha$ for different aperture distributions.

| Aperture | Overall efficiency (%) | $\alpha$ |
|---|---|---|
| H: -90° to 90° | 34.2 | 1.8 |
| I: -60° to 60° | 18.5 | 3.2 |

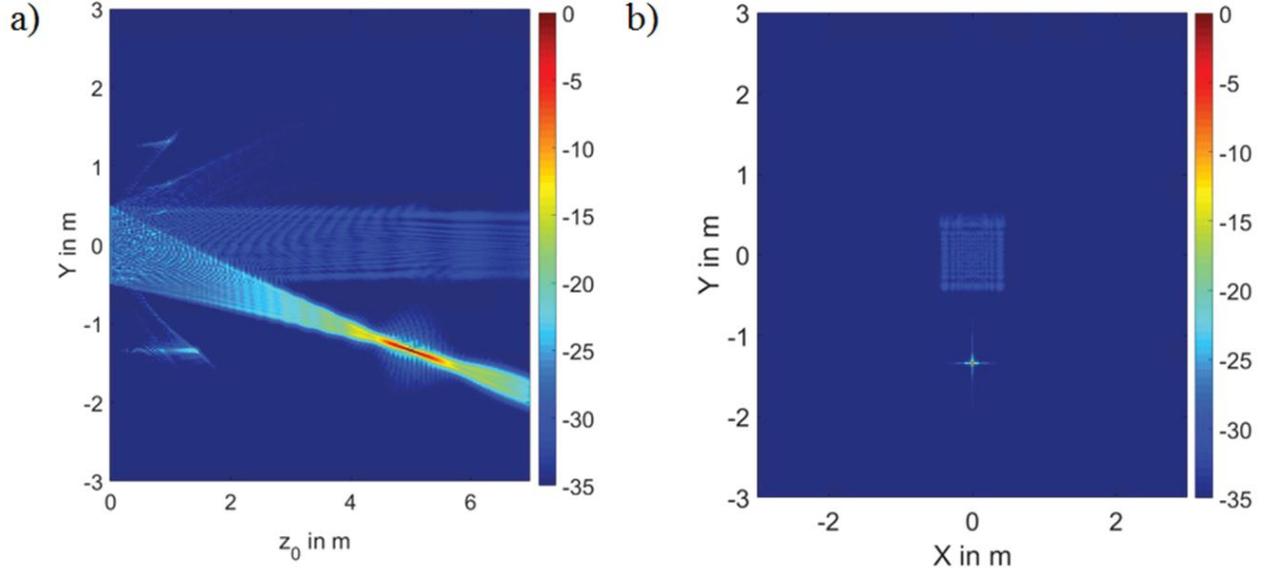

FIG. 11. Field distributions of a Metasurface designed to produce a focus at 15° away from the normal. (a) Intensity distribution in dB of the y-z plane showing higher diffraction orders (b) Intensity distribution in dB of the x-y plane (cross range) plot of off axis focusing with the zeroth order mode at the center.

### E. Sampling Criteria

The number of points at which the aperture plane is sampled is a parameter worthy of consideration. Each sampling point, or pixel, represents a point that requires dynamic control, necessitating a radiating element with tuning mechanism and an integrated bias/control circuit. If we consider $\lambda_0/8$ sampling for an area of 1 x 1 m², several millions of metasurface elements and associated circuitry would be necessary. We can arrive at an estimate of the required sampling by considering the analytic form of the fringes over the aperture at their most extreme spatial variation.

We consider the two-dimensional case (fields and sources invariant along the x-direction) to simplify the math. The interference, or fringe pattern, on the aperture from a line source located at $(y_0, z_0)$ has the form:

$$\cos\left(k\sqrt{(y-y_0)^2 + z_0^2}\right), \tag{15}$$

We are interested in the spatial variation at the most extreme portion of the interference pattern, which is the farthest away from the central spot. For the off-axis hologram, the most rapid variation will occur at the edge of the aperture, or at:

$$y = -\frac{D}{2} + \Delta y, \tag{16}$$

where $\Delta y$ is the distance away from point at the edge of the aperture. Substituting Eq. 16 into Eq. 15:

$$\cos\left(k\sqrt{\left(\Delta y - \frac{D}{2} - y_0\right)^2 + z_0^2}\right), \qquad (17)$$

Since $\Delta y$ is smaller than any of the other quantities, we can take a small argument expansion to obtain

$$\cos\left(kz_0\left(1 + \frac{\left(\frac{D}{2} + y_0\right)}{z_0^2}\Delta y\right)\right), \qquad (18)$$

A full cycle for one fringe occurs over a distance of

$$\Delta y = \frac{\lambda_0 z_0}{\frac{D}{2} + y_0}. \qquad (19)$$

Using Eq. 19 with $z_0 = 1$m, $\lambda_0 = 0.0039$ m, $D = 1$m and $y_0 = 1.34$ m, we obtain $\Delta y = 0.0021$ m, which is fairly close to $\lambda_0/2$ (0.0019 m). This analysis suggests that we can sample the aperture at a spacing of $\lambda_0/2$, reducing the number of metasurface elements by a factor of 16. A focus near the aperture and at a fairly significant angle from the normal was chosen, since the spatial variation of the fringe pattern is most rapid for such focal spots.

## IV. MICROWAVE AND MILLIMETER WAVE SOURCES AND ENERGY HARVESTERS

The development of low cost microwave and millimeter wave sources and detectors has been driven by the growth of high volume markets in wireless communication and automotive radar. Unlike traditional defense markets, where performance is the primary criterion, cost sensitive consumer markets drive the development of standard components that may be useful in beamed WPT applications. In particular, the coming generation of 5G wireless networks will depend on low cost, highly integrated silicon RF solutions. Silicon (CMOS) RF integrated circuits (RFICs) have been demonstrated well into the millimeter wave bands using the latest generation of sub-45nm CMOS process technologies, where transistor FT can exceed 200 GHz[29,30]. Leveraging scaled CMOS processes, millimeter wave RF synthesizers and low noise amplifiers can be integrated with complex digital control systems on the same semiconductor die.

One limitation of CMOS millimeter wave technology is the limited power handling capability of silicon CMOS FETs. 45nm CMOS transistors typically have a drain breakdown voltage limit of approximately 1.1V and a peak drain power density of 100 mW/mm. Coupled with the relatively low thermal conductivity of silicon (149 W/m-K), these limitations suggest that alternative semiconductor technologies have an important role to play in WPT sources. The current generation of gallium nitride (GaN) on silicon carbide (SiC) high electron mobility transistors (HEMTs) excel from the perspective of high drain voltage and excellent power density. Current GaN HEMT technology supports a power density of up to 5 W/mm, over 50X greater than silicon CMOS devices. GaN power amplifiers have been demonstrated at power levels over 100 W in the K-band (17-26 GHz)[31], albeit with power added efficiency of only 25-30%. Future GaN devices with optimized geometry, configured for narrowband WPT applications, will likely improve significantly on this starting point. It should be noted that vacuum electronics, such as gyrotron tubes, can be very efficient in the high power regime (up to 100 kW or more), but their reliability and bulky high voltage power supply requirements likely render vacuum electronics unsuitable for consumer applications.

Integrated antennas with RF energy harvesters (rectennas) based on low cost Schottky diodes are a well-established area of research[18]. With careful attention to packaging, Schottky diodes are useful up to 0.1 THz or beyond. Schottky diodes integrated with rectenna elements have been demonstrated using on-chip antenna systems that enable a small size and relatively high efficiency - for example, efficiency of 53% at 35 GHz and 37% at 94 GHz have been reported using on-chip rectennas[32].

**V. CONCLUSION**

A beamed wireless power transfer (WPT) system is not without challenges, but presents an interesting alternative to near-field magnetic coupling schemes. In particular, the possibility of selectively beaming power to small devices located anywhere within a volume is a desirable advantage. Emerging beam-steering technologies, such as the metasurface aperture analyzed here, have the potential to reach the low price points required for larger-scale adoption in consumer driven markets. The initial studies presented here provide some guidance relating aperture size and wavelength to system parameters such as transfer efficiency and coverage area.

For the studies pursued here, an ideal holographic metasurface aperture with plane wave illumination was considered. A practical system would likely make use of a guided mode rather than a free space wave to create a low-profile device as pictured in Fig. 1. For such guided wave implementations, higher order diffracted beams can be further rejected because of the natural phase shift of the reference wave over the aperture[33].

The end-to-end efficiency of a Fresnel-zone WPT system depends on three major factors: the RF source efficiency, the aperture and coupling efficiency analyzed here, and the RF-to-DC conversion efficiency of the receiving energy harvester. At K-band, current efficiency for commercially available solid state sources (~ 30%) and energy harvesters (~ 53%) limits end-to-end efficiency to around 15%. Assuming an 80% efficient aperture, end-to-end efficiency of around 10-12% seems feasible with current technology. While the efficiencies of the constrained apertures considered here may be lower than would be desirable for a commercial system, the phase-, amplitude- and Lorentzian-constrained holograms provide basic trends and suggest an important area of future research. By carefully optimizing the aperture to refine the phase/amplitude distributions, considerable improvement can be made to the overall efficiency and reduction of unwanted focal spots and diffracted beams.

While we have presented simulations for one off-axis focusing scenario, we expect that the trends found for the efficiencies and other metrics are representative for all focal spots, because the beam waists of the focal points over the entire simulation domain (for holograms with different focal lengths) do not change with limitations to the phase or amplitude distribution, and are very close to that predicted by Gaussian optics. The degradation in efficiency due to diffracted orders is roughly similar for focal points at any position.

## ACKNOWLEDGMENTS

This work was supported by the Air Force Office of Scientific Research (AFOSR, Grant No. FA9550-12-1-0491).

## APPENDIX: EFFECTIVE APERTURE SIZE FOR OFF AXIS FOCUS

Since we are creating an ideal hologram over the aperture for every focal position in the range of coverage, there are no aberrations introduced to the focus. The widening of the beam waist must arise entirely from the loss of aperture, which we can estimate from the geometry. Consider the situation depicted in Fig. 12.

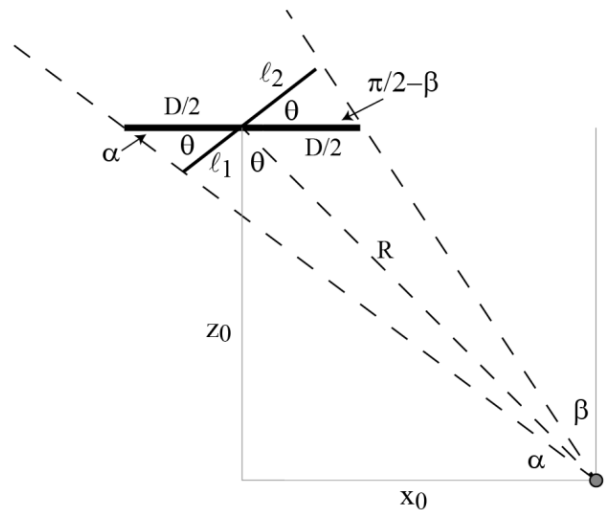

FIG. 12. Illustration of the effective aperture considered for the analytical calculation of the beam waist for off-axis focusing.

The thicker solid line represents the actual aperture. However, we can conceptually replace this aperture, which makes an angle $\theta$ with respect to the position of the focal spot, by a second aperture, represented by the thinner line, for which the focal spot is now on-axis. The characteristics of the focal spot must be the same for either aperture, since we assume the hologram is ideal for both cases. Given the position of the focal spot, we have the following relationships:

$$\cos\theta = \frac{z_0}{\sqrt{x_0^2 + z_0^2}}$$

$$\tan\alpha = \frac{z_0}{x_0 + \frac{D}{2}}, \tag{A1}$$

$$\tan\beta = \frac{x_0 - \frac{D}{2}}{z_0}$$

And $R = \sqrt{x_0^2 + z_0^2}$. We seek the length of the effective aperture. Designating the lengths $l_1$ and $l_2$ for the two sections (on either side) of the effective aperture, we can apply the law of sines as follows:

$$\frac{\sin\alpha}{l_1} = \frac{\sin(\alpha - \theta)}{D/2}$$

$$\frac{\sin\left(\frac{\pi}{2} - \beta\right)}{l_2} = \frac{\sin\left(\frac{\pi}{2} + \beta - \theta\right)}{D/2}, \tag{A2}$$

which yields

$$l_1 = \frac{D}{2}\frac{\sin\alpha}{\sin(\alpha + \theta)}$$

$$l_2 = \frac{D}{2}\frac{\cos\beta}{\cos(\beta - \theta)}, \tag{A3}$$

Thus,

$$D_{eff} = l_1 + l_2 = \frac{D}{2}\left[\frac{1}{\cos\theta + \sin\theta\cot\alpha} + \frac{1}{\cos\theta + \sin\theta\tan\beta}\right], \tag{A4}$$

While this formula for the effective aperture is not particularly illuminating, we can rearrange the formula to find, after some algebra,

$$D_{eff} = Dz_0\left[\frac{R^3}{R^4 + \frac{D^2 x_0^2}{4}}\right], \tag{A5}$$

or,

$$D_{eff} = D\cos\theta \left[\frac{1}{1+\frac{1}{4}\left(\frac{D}{R}\right)^2 \sin^2\theta}\right], \tag{A6}$$

Eq. A6 shows that away from the aperture, where R>>D, the aperture reduction goes simply as the cosine of the angle between the aperture axis and the focal position. Closer to the aperture, however, the aperture reduction occurs more quickly. For the off-axis Gaussian beam waist, then, we should use the effective aperture. Note that for the effective aperture, the focal length $z_0$ is equal to R, so that

$$w_0 = \frac{4}{\pi}\frac{\lambda_0}{D}z_0 = \frac{4}{\pi}\frac{\lambda_0}{D_{eff}}R = \frac{4}{\pi}\frac{\lambda_0}{D_{eff}}\frac{R}{z_0}z_0 = \frac{4}{\pi}\frac{\lambda_0}{D_{eff}\cos\theta}z_0, \tag{A7}$$

or,

$$w_0 = \frac{4}{\pi}\frac{\lambda_0 z_0}{D\cos^2\theta}\left[1+\frac{1}{4}\left(\frac{D}{R}\right)^2 \sin^2\theta\right], \tag{A8}$$

If we can neglect the term in brackets—a good approximation for the cases under consideration—we obtain a simple formula for the beam waist valid for off-axis focusing:

$$w_0 = \frac{4}{\pi}\frac{\lambda_0 z_0}{D\cos^2\theta} . \tag{A9}$$

as used in the text above.